\documentclass[pra,aps,10pt,twocolumn,superscriptaddress,showpacs,floatfix,longbibliography,citeautoscript]{revtex4-1}
\usepackage{graphicx}
\usepackage{color}
\usepackage{latexsym}
\usepackage{amsmath}
\usepackage{amsfonts}
\usepackage{amssymb}
\usepackage{subfigure}
\usepackage{bm}
\usepackage{float}
\usepackage{tabularx}

\usepackage{hyperref}
\hypersetup{colorlinks=true}
\usepackage[all]{hypcap} % let hyperlinks correctly point to figures rather than their captions; must be loaded after the hyperref package!

\begin{document}

\title{
Nonlinear interaction effects in a three-mode cavity optomechanical system
}

\author{Jing Qiu}
%\email{li-jing.jin@lpmmc.cnrs.fr}
\affiliation{Beijing Computational Science Research Center, Beijing 100193, China}

\author{Li-Jing Jin}
%\email{li-jing.jin@lpmmc.cnrs.fr}
\affiliation{Institute for Quantum Computing, Baidu Research, Beijing 100193, China}
\affiliation{Beijing Computational Science Research Center, Beijing 100193, China}

\author{Stefano Chesi}
\email{stefano.chesi@csrc.ac.cn}
\affiliation{Beijing Computational Science Research Center, Beijing 100193, China}
\affiliation{Department of Physics, Beijing Normal University, Beijing 100875, China}

\author{Ying-Dan Wang}
\email{yingdan.wang@itp.ac.cn}
\affiliation{CAS Key Laboratory of Theoretical Physics, Institute of Theoretical Physics, Chinese Academy of Sciences, P.O. Box 2735, Beijing 100190, China}
\affiliation{School of Physical Sciences, University of Chinese Academy of Sciences, No.19A Yuquan Road, Beijing 100049, China}
\affiliation{Synergetic Innovation Center for Quantum Effects and Applications, Hunan Normal University, Changsha 410081, China}

\begin{abstract}

We investigate the resonant enhancement of nonlinear interactions in a three-mode cavity optomechanical system with two mechanical oscillators. By using the Keldysh Green's function technique we find that nonlinear effects on the cavity density of states can be greatly enhanced by the resonant scattering of two phononic polaritons, due to their small effective dissipation. In the large detuning limit and taking into account an upper bound on the achievable dressed coupling, the optimal point for probing the nonlinear effect is obtained, showing that such three-mode system can exhibit prominent nonlinear features also for relatively small values of $g/\kappa$.

\end{abstract}

\date{\today}

\maketitle

\section{Introduction}

Optomechanical systems \cite{Aspelmeyer2014RMP} have witnessed remarkable progress in controlling the quantum state of the coupled photonic and mechanical modes. Some highlights are the demonstration of mechanical ground-state cooling \cite{teufel2011sideband,chan2011laser,clark2017sideband}, generation of strongly squeezed light \cite{safavi2013squeezed,purdy2013strong} and mechanical states \cite{wollman2015quantum,PhysRevLett.117.100801}, coherent transduction \cite{andrews2014bidirectional}, and entanglement of remote mechanical oscillators \cite{riedinger2018remote,ockeloen2018stabilized}. All these applications are based on a linearized interaction under strong optical drive, when the optomechanical coupling is greatly enhanced by the large number of intracavity photons. Continuous technical progress has allowed the dressed coupling $G$ to enter and even surpass the strong-coupling regime \cite{groblacher2009observation,Teufel2011Nature,Verhagen2012Nature,Peterson2019PRL}.

On the other hand, nonlinear interactions are necessary for the generation of non-classical states and a variety of interesting effects were predicted \cite{rabl2011photon, Komar2013PRA, Ludwig2012PRL, Liu2013PRL, Xu2015PRA, Borkje2013PRL, Lemonde2013PRL, Lemonde2015PRA, Lemonde2016NC, Jin2018PRA}. For these nonlinear signatures the relevant energy scale is the single-photon optomechanical coupling $g$ which, unfortunately, remains much smaller than both the mechanical frequency $\omega_m$ and cavity damping $\kappa$ in virtually all setups with solid-state oscillators.

Some proposals for effectively enhancing the single-photon coupling strength consider modifying the type of drive, e.g., by introducing a squeezed optical input or a mechanical parametric drive \cite{Lemonde2016NC,lu2015squeezed,yin2017nonlinear}. Recently, it was also shown that multi-mode setups can allow for a large enhancement of nonlinear effects \cite{Jin2018PRA}. An attractive feature of the latter scheme is that relies on an optomechanical chain which is very close to existing experimental setups. In particular, it is essentially equivalent to four-mode optomechanical systems developed for efficient nonreciprocity \cite{peterson2017demonstration, bernier2017nonreciprocal}.

The aim of the present work is to explore if the optomechanical chain of Ref.~\cite{Jin2018PRA} can be further simplified, while preserving a large enhancement factor of the nonlinear signatures with respect to the two-mode system. In an optomechanical cavity, the largest nonlinear effects on the optical density of states (DOS) are due to a resonant scattering process between polaritons (i.e., the coupled eigenmodes of the linearized system)~\cite{Borkje2013PRL, Lemonde2013PRL, Lemonde2015PRA}. The main advantage of multi-mode setups is that two of these polaritons (instead of one) can be mechanical modes weakly hybridized with the optical cavities. Although the nonlinear coupling between phonon-like polaritons is much smaller than $g$, the reduction of the scattering amplitude is compensated by the exceptional coherence properties of the polaritons, whose lifetime is only limited by the mechanical damping $\gamma \ll \kappa$ \cite{Jin2018PRA}.

\begin{figure}
\begin{center}
\includegraphics[width=0.45\textwidth]{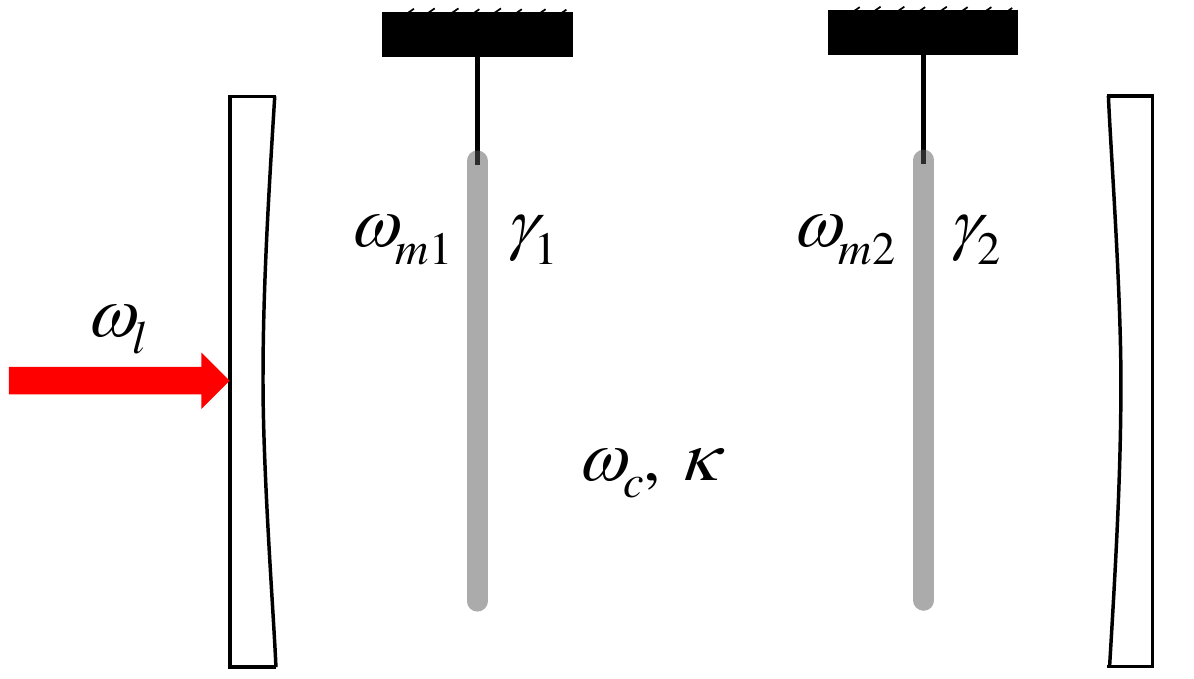}
\end{center}
\caption{Schematic illustration of the three-mode optomechanical systems.
$\omega_{c}$ is the cavity frequency and $\kappa$ is the cavity damping rate.
$\omega_{mi}$ are the mechanical frequencies and $\gamma_{i}$ are the mechanical damping rates.
$\omega_{l}$ is the frequency of the laser drive, represented by the red arrow.
\label{fig:Three_mode_cavity}}
\end{figure}

The above discussion makes intuitively clear that a single cavity interacting with two mechanical oscillators  (see Fig.~\ref{fig:Three_mode_cavity}) is the minimal setup where this physics can take place. Indeed, we find that in a three-mode setup the typical figure of merit $(g/\kappa)^2$ of nonlinear effects can be enhanced by a large factor which, quite naturally, depends on the ratio $\kappa/\gamma$. The enhancement is large also far from the optomechanical instability, and can be optimized with respect to the ratio of the two mechanical frequencies. Doing so, we find that the largest enhancement is proportional to $(G/\omega_{m})^2$, thus is particularly interesting in view of the recent success in achieving the ultra-strong coupling regime \cite{Peterson2019PRL}.

The outline of our paper is as follows: In Sec.~\ref{Sec: the system} we introduce the model and in Sec.~\ref{Sec: Polariton eigenmodes} we diagonalize the linear part of the Hamiltonian. The approach to include nonlinear effects is described in Sec.~\ref{sec:method} and applied numerically in Sec.~\ref{Sec: Resonant enhancement}. Physical understanding of the results, together with an approximate analytical treatment in the most relevant regime of large detuning, is provided in Sec.~\ref{Sec:Result_2}. Finally, we conclude in Sec~\ref{Sec: summary} and give some technical details in Appendices~\ref{appendix_equal_wm} and \ref{Appendix_V}.

%%%%%%%%%%%%%%%%%%%%%%%%%%%%%%%%%%%%%%%%%%%%%%%%%%%%%%%%%%%%%%%
%%%%%%%%%%%%%%%%%%%%%%%%%%%%%%%%%%%%%%%%%%%%%%%%%%%%%%%%%%%

%%%%%%%%%%%%%%%%%%%%%%%%%%%%%%%%%%%%%%%%%%%%%%%%%%%%%%%%%%%%%%%
\section{Model}\label{Sec: the system}
%%%%%%%%%%%%%%%%%%%%%%%%%%%%%%%%%%%%%%%%%%%%%%%%%%%%%%%%%%%%%%%

As shown in Fig.~\ref{fig:Three_mode_cavity}, we consider a driven optomechanical cavity with two mechanical oscillators. The system is described by the Hamiltonian
$H=H_{0}+H_{\mathrm{diss}}$, with:
\begin{align} \label{H0}
H_{0}  = & \omega_{c}a^{\dagger}a+ \left(\alpha e^{-i\omega_{l}t}a^{\dagger}+\mathrm{H.c.} \right) \nonumber \\
 & +\sum_{i=1,2}\left( \omega_{mi}b_{i}^{\dagger}b_{i} +g_{i}a^{\dagger}a(b_{i}+b_{i}^{\dagger})\right).
\end{align}
Here, $a$ is the annihilation operator for cavity mode, $\omega_{c}$
is the cavity frequency, $b_{i}$ $(i=1,2)$ are the annihilation operators
of the mechanical modes, $\omega_{mi}$ are the mechanical frequencies (we take $\omega_{m2} \geq \omega_{m1}$),
$g_{i}$ is the single-photon optomechanical
coupling, and $\alpha$ is proportional to the amplitude of a classical
drive at frequency $\omega_{l}$. $H_{\mathrm{diss}}$ describes the
dissipation of photons and phonons by independent baths:
\begin{align}
H_{\mathrm{diss}}=& \sum_{i=1,2}\sum_{j}\omega_{mi,j}f_{mi,j}^{\dagger}f_{mi,j} + \sum_{j}\omega_{c,j}f_{c,j}^{\dagger}f_{c,j} \nonumber \\
&-i\sum_{i=1,2}\sum_{j}\sqrt{\frac{\gamma_{i}}{2\pi\rho_{mi}}}(f_{mi,j}-f_{mi,j}^{\dagger})(b_{i}+b_{i}^{\dagger}) \nonumber \\
&-i\sum_{j}\sqrt{\frac{\kappa}{2\pi\rho_{c}}}(f_{c,j}-f_{c,j}^{\dagger})(a+a^{\dagger}),
\label{eq:dissipation term}
\end{align}
where $f_{c,j}$ is the annihilation operator for cavity-bath mode $j$
(frequency $\omega_{c,j}$), $\kappa$ is the damping rate of photons
inside the cavity, and $\rho_{c}$ is the cavity-bath density of states.
Furthermore, $f_{mi,j}$ is the annihilation operator for bath mode $j$ of mechanical resonator $i=1,2$
(frequency $\omega_{mi,j}$), $\gamma_{i}$ are the two mechanical damping rates, and $\rho_{mi}$ are the mechanical density of states.
Because we consider a Markovian bath, we take $\kappa$, $\rho_{c}$,  $\gamma_{i}$, and $\rho_{mi}$ to be frequency independent.

After transforming the cavity mode to a frame rotating at the laser frequency $\omega_{l}$, and performing a standard displacement transformation $a=\overline{a}+d$ (where $\overline{a}$ is the classical cavity amplitude induced by the
laser drive), the Hamiltonian of the system takes the form $H_{l}+H_{nl}$ where
\begin{equation}
H_{l}  =  -\Delta d^{\dagger}d+\sum_{i=1,2}\left(\omega_{mi}b_{i}^{\dagger}b_{i}+G_{i}(d+d^{\dagger})(b_{i}+b_{i}^{\dagger})\right).
\label{eq:linear term}
\end {equation}
Here, $\Delta=\omega_{l}-\omega_{c}$ is the detuning and
$G_{1,2}=g_{1,2}\overline{a}$ are the dressed couplings which, for definiteness, we take as real. The average number of photons in the cavity is $N=\bar{a}^2$. While $H_{l}$ describes the linear interactions between photon
modes and mechanical modes, $H_{nl}$ includes the intrinsically
nonlinear interaction of system:
\begin{equation}\label{eq:nonlinear term}
H_{nl}  =  g_{1}d^{\dagger}d(b_{1}+b_{1}^{\dagger})+g_{2}d^{\dagger}d(b_{2}+b_{2}^{\dagger}).
\end{equation}
Throughout this work we focus on a red-detuned laser (i.e., $\Delta<0$), which allows to avoid optomechanical instabilities in large range of parameters. Finally, $H_{\mathrm{diss}}$ is transformed in a similar way. In a frame rotating at $\omega_l$ for the phonon bath modes and after a rotating-wave approximation, the final form is similar to Eq.~(\ref{eq:dissipation term}) except for the replacements $\omega_{c,j} \to \Delta_{c,j} = \omega_{c,j} -\omega_l $ and $(f_{c,j}-f_{c,j}^{\dagger})(a+a^{\dagger}) \to (f_{c,j}d^\dag-f_{c,j}^{\dagger}d)$.

%%%%%%%%%%%%%%%%%%%%%%%%%%%%%%%%%%%%%%%%%%%%%%%%%%%%%%%%%%%%%
\section{Polariton eigenmodes}\label{Sec: Polariton eigenmodes}
%%%%%%%%%%%%%%%%%%%%%%%%%%%%%%%%%%%%%%%%%%%%%%%%%%%%%%%%%%%%%

As a first step, we consider the diagonalization of the linear problem via a Bogoliubov transformation  (where $T$ indicates the transpose):
\begin{equation}
\left(\begin{array}{ccc}
b_{1} & b_{2} & d
\end{array}\right) ^T  =   V\left(\begin{array}{cccccc}
c_{1} & c_{2} & c_{3} & c_{1}^{\dagger} & c_{2}^{\dagger} & c_{3}^{\dagger}\end{array}\right)^{T},  \label{eq:trans_relat}
\end{equation}
leading to $H_{l}=\sum_{i=1,2,3}\omega_{i}c_{i}^{\dagger}c_{i}$. We impose
\begin{equation}
\omega_{3}\geq\omega_{2}\geq\omega_{1} \geq 0,\label{eq:frequencies_relation}
\end{equation}
where the requirement of positive frequencies is to ensure the stability of linear problem (this condition neglects the effect of small damping rates $\kappa,\gamma_i$). The $c_i$ are polariton modes, given by linear combinations of the cavity and mechanical modes. The matrix $V$ can be most easily found in the coordinate representation, in which the quadratures $x_i, p_i$ ($i=1,2,3$) are defined through $b_i =  (\omega_{mi}x_i+ip_i)/\sqrt{2\omega_{mi}}$ for $i=1,2$ and $d =  (|\Delta|x_3+ip_3)/\sqrt{2|\Delta|}$. With this notation, the linear Hamiltonian $H_l$ takes the form:
\begin{equation}
H_l = \sum_{i}\frac{p_i^2}{2} + \frac12 \sum_{i,j} x_i M_{i,j} x_j ,
\end{equation}
where
\begin{align} \label{M_matrix}
M & =\left(\begin{array}{ccc}
\omega_{m1}^{2} &  0 & 2G_{1}\sqrt{\left|\Delta\right|\omega_{m1}} \\
0 & \omega_{m2}^{2} & 2G_{2}\sqrt{\left|\Delta\right|\omega_{m2}}  \\
 2G_{1}\sqrt{\left|\Delta\right|\omega_{m1}} & 2G_{2}\sqrt{\left|\Delta\right|\omega_{m2}}& \Delta^{2}
\end{array}\right).
\end{align}
$M$ is diagonalized by a (orthogonal) matrix $U$. Explicitly, $(U^T MU)_{i,j} = \omega_i^2 \delta_{i,j}$. Then $V$ can be written in block-matrix form:
\begin{equation}\label{V_blocks}
V = (V_+ \,  V_-),
\end{equation}
where $V_\pm$ are related to $U$ as follows:
\begin{equation}\label{Vpm}
V_\pm =  \left(\begin{array}{ccc}
U_{11}f_\pm \left(\frac{\omega_{m1}}{\omega_{1}}\right) & U_{12}f_\pm \left(\frac{\omega_{m1}}{\omega_{2}}\right)  & U_{13}f_\pm \left(\frac{\omega_{m1}}{\omega_{3}}\right)  \\
U_{21}f_\pm \left(\frac{\omega_{m2}}{\omega_{1}}\right) & U_{22}f_\pm \left(\frac{\omega_{m2}}{\omega_{2}}\right)  & U_{23}f_\pm \left(\frac{\omega_{m2}}{\omega_{3}}\right)  \\
U_{31}f_\pm \left(\frac{\left|\Delta\right|}{\omega_{1}}\right) & U_{32}f_\pm \left(\frac{\left|\Delta\right|}{\omega_{2}}\right)  & U_{33}f_\pm \left(\frac{\left|\Delta\right|}{\omega_{3}}\right)
\end{array}\right),
\end{equation}
with $f_\pm (x)=(\sqrt{x}\pm \sqrt{1/x})/2 $. Clearly, all matrix elements of $V$ are real.

Unfortunately, analytic expressions of $U$ and $V$ are not available in general. This is  at variance with the optomechanical ring treated in Ref.~\cite{Jin2018PRA}, where translational invariance allows to transform the multi-mode problem into independent 2-mode systems. Here, only the special case $\omega_{m1}=\omega_{m2}$ can be easily treated as a 2-mode system, by introducing a `dark' and `bright' mechanical mode as discussed in Appendix~\ref{appendix_equal_wm}. Although $M$ is easily diagonalized, the dark mode is completely decoupled from the cavity (even after including the nonlinear interaction) and scattering between phonon-like modes is not allowed.

Since the nonlinear effects at  $\omega_{m1}=\omega_{m2}$ are of the same order of a simple optomechanical cavity, we should consider the general case $\omega_{m1}\neq \omega_{m2}$. We will be able to obtain analytical expressions in the relevant regime of large detuning, based on a perturbative treatment. For this approach we require $|\Delta| \gg \omega_{mi},G_i$, to have the off-diagonal elements of $M$ smaller than the gap  $\Delta^2$, see  Eq.~(\ref{M_matrix}). Within this approach we obtain the eigenfrequencies as follows (see Appendix~\ref{Appendix_V}):
\begin{align}\label{w12}
\omega^2_{1,2}  \simeq  & \frac12 \bigg(   \omega_{m1}^{2}+ \omega_{m2}^{2}- B^2_{11}- B^2_{22}  \nonumber \\
& \mp \sqrt{\left(\omega_{m1}^{2}- \omega_{m2}^{2}-B^2_{11}+ B^2_{22}\right)^2 + 4B_{12}^{4}} \bigg), \\
\omega^2_{3}    \simeq & \Delta^2 + B^2_{11}+ B^2_{22} ,
\end{align}
where we defined
\begin{equation}\label{B_def}
B^2_{ij}  =  \frac{4}{|\Delta|}G_{i}G_j \sqrt{\omega_{mi}\omega_{mj}}.
\end{equation}
The sign in Eq.~(\ref{w12}) is chosen to satisfy $\omega_2 \geq \omega_1$. Using $\omega_1^2 \geq 0$ we get the stability condition:
\begin{equation}\label{eq:stability}
4G_{1}^{2}\omega_{m2}+4G_{2}^{2}\omega_{m1}\leq\left|\Delta\right|\omega_{m1}\omega_{m2}.
\end{equation}

%%%%%%%%%%%%%%%%%%%%%%%%%%%%%%%%%%%%%%%%%%%%%%%%%%%%%%%%%%%%%%%
\section{General formalism}\label{sec:method}
%%%%%%%%%%%%%%%%%%%%%%%%%%%%%%%%%%%%%%%%%%%%%%%%%%%%%%%%%%%%%%%

To characterize the effects of the nonlinear interaction, we follow the treatment developed in Ref.~\cite{Lemonde2013PRL} for the two-mode system and extended to multi-mode optomechanical chains in Ref.~\cite{Jin2018PRA}. Within this approach, the retarded photon Green's function $ G^{R}[d,d^{\dagger};\omega]=-i\int_{-\infty}^{+\infty}dte^{i\omega t}\left\langle  [d(t),d^{\dagger}(0) ]\right\rangle \theta(t)$ (where $\theta(t)$ is the Heaviside step function) is computed with the Keldysh diagrammatic technique by including the nonlinear interaction $H_{nl}$ through a dominant second-order correction to the self-energy. This approach is justified by the smallness of the nonlinear interaction.

Several observable quantities can be extracted from $G^{R}[d,d^{\dagger};\omega]$ and we will focus on the cavity DOS $\rho_{d}(\omega)$, defined as follows:
\begin{equation}\label{eq:DOS}
\rho_{d}(\omega)=-\frac{1}{\pi}\mathrm{Im}G^{R}[d,d^{\dagger};\omega].
\end{equation}
The modification of the optomechanically induced transparency (OMIT) signal can also be easily extracted from $G^R[d,d^{\dagger};\omega]$, and is directly related to $\rho_d(\omega)$. The relation between the polaritons and photon Green's functions immediately follows from Eq.~(\ref{eq:trans_relat}), leading to the following expression of $\rho_d$ in terms of the polariton retarded Green's functions:
\begin{align}\label{eq:rho_d}
\rho_{d}(\omega)  =  & -\frac{1}{\pi}{\rm Im}\bigg\{ \sum_{i=1}^3 \left( V_{3,i}^{2}   G^{R}[c_i,c_i^{\dagger};\omega] \right.   \nonumber \\
& \left. + V_{3,i+3}^{2}G^{R}[c_i^{\dagger},c_i;\omega]\right) \bigg\},
\end{align}
where we have neglected the contribution from the small off-diagonal components $ G^{R}[c_i,c_j^{\dagger};\omega] $, with  $ i \neq j$ (which is justified where the nonlinear interaction and polariton dampings are much smaller than the differences between polariton frequencies). Considering the nonlinear interaction, the Green's functions entering Eq.~(\ref{eq:rho_d}) are:
\begin{equation}\label{GR}
G^{R}[c_i,c_i^{\dagger};\omega]=\frac{1}{\omega-\omega_{i}+i\frac{\kappa_{i}}{2}-\Sigma_i^{R}(\omega)},
\end{equation}
and $G^{R}[c_i^{\dagger},c_i;\omega]=(G^{R}[c_i,c_i^{\dagger};-\omega])^*$. In Eq.~(\ref{GR}), $\kappa_{i}$ is the effective dissipation of polariton $i$ and $\Sigma_i^{R}(\omega)$ is the retarded self-energy. The explicit form of these quantities is discussed below.

First we consider the effect of $H_{\rm diss}$, giving the following damping rates of the polaritons:
\begin{equation}\label{kappa_i}
\kappa_{i}  =   \kappa V_{3,i}^{2}-\kappa V_{3,i+3}^{2}+\sum_{j=1,2}\gamma_{j}\left(V_{j,i}+V_{j,i+3}\right)^{2},
\end{equation}
where we can recognize three distinct contributions (from the cavity and two phonon baths). The different form in which the matrix elements $V_{i,j}$ enter the photon- and photon-bath contributions is due to the quantum heating induced by the drive on the cavity mode, leading to bath modes with negative frequency in the rotating frame ($\Delta_{c,i}>-\omega_l$). The unperturbed retarded Green's function is simply given by $ G_{0}^{R}[c_i,c_i^{\dagger};\omega]=1/\left(\omega-\omega_{i}+i\kappa_{i}/2\right)$. In the perturbative calculation of $\Sigma_i^{R}(\omega)$, it is also necessary to consider the Keldysh Green's function, $ G_{0}^{K}[c_i,c_i^{\dagger};\omega]= 2i(2n_i+1){\rm Im}G_{0}^{R}[c_i,c_i^{\dagger};\omega]$. Here $n_i$ are the occupation numbers of the free polaritons, which can also be found from $H_{\rm diss}$:
\begin{equation} \label{n_i}
n_{i}  = \frac{\kappa}{\kappa_{i}} V_{3,i+3}^{2}+\sum_{j=1,2}\frac{\gamma_{j}}{\kappa_{i}}\left(V_{j,i}+V_{j,i+3}\right)^{2}n_{B}(\omega_{i}),
\end{equation}
where $n_{B}(\omega_{i})=1/(e^{\beta\omega_i}-1)$ is the Bose-Einstein distribution function, evaluated at the frequency of polariton $c_{i}$ and the (physical) temperature of the two mechanical baths. In the above expression we assume the optical cavity bath to be effectively at zero temperature.

Now we consider the self-energy induced by the nonlinear interaction Eq.~(\ref{eq:nonlinear term}), which is useful to rewrite in the polariton basis as:
\begin{align}\label{H_nl_4terms}
H_{nl} = & \big( g_{322} c_{3}c_{2}^{\dagger}c_{2}^{\dagger} + g_{311} c_{3}c_{1}^{\dagger}c_{1}^{\dagger}+ g_{321} c_{3}c_{2}^{\dagger}c_{1}^{\dagger}  \nonumber \\
& + g_{211} c_{2}c_{1}^{\dagger}c_{1}^{\dagger} + {\rm H.c.} \big) + \ldots,
\end{align}
where we rely on the fact that each nonlinear term can only contribute appreciably if the system is close to a resonant condition. Therefore, we dropped contributions which obviously cannot be resonant, e.g., terms $\propto c_{i}^{\dagger}c_{j}^{\dagger}c_{k}^{\dagger}$.  Among the terms $\propto c_{i}c_{j}^{\dagger}c_{k}^{\dagger}$, the only important ones are the four explicitly written in Eq.~(\ref{H_nl_4terms}), after taking into account the requirement $\omega_ i \simeq \omega_j + \omega_k$ and our conventional ordering of eigenfrequencies~(\ref{eq:frequencies_relation}).

The various contributions to the self-energy arising from Eq.~(\ref{H_nl_4terms}) can be computed in a relatively straightforward way following the discussion of 2-mode and 4-mode systems \cite{Lemonde2013PRL, Jin2018PRA}. However, as a further simplification, we will consider the regime of low-energy polaritons $i=1,2$ with predominantly mechanical character. In this case, it is expected that the effect of scattering process $g_{211} c_{2}c_{1}^{\dagger}c_{1}^{\dagger} $ is greatly enhanced, due to the long lifetime of the polariton modes \cite{Jin2018PRA}. From Eq.~(\ref{M_matrix}) we see that a simple condition to quench the mixing of optical and mechanical modes is $|\Delta| \gg \omega_{mi},G_i$, i.e., the perturbative regime mentioned already. Since  $\omega_3 \simeq |\Delta| \gg \omega_{1,2}$, it is quite clear that the condition $\omega_3 = \omega_j +\omega_k$ cannot be realized, which justifies neglecting the first line of Eq.~(\ref{H_nl_4terms}).

In summary, in the rest of the paper we will focus on a parameter regime where the nonlinear interaction can be approximated as:
\begin{equation}\label{H_nl_211}
H_{nl} \simeq  g_{211} ( c_{2}c_{1}^{\dagger}c_{1}^{\dagger}+ c_{1}c_{1} c_{2}^\dagger ).
\end{equation}
The explicit expression of $g_{211}$ is
\begin{align}
g_{211}  = \sum_{i=1,2} g_{i}& \left[\left(V_{3,1}V_{3,2}+V_{3,4}V_{3,5}\right)\left(V_{i ,1}+V_{i, 4}\right)\right.\nonumber \\
   & \left.+V_{3,1}V_{3,4}\left(V_{i,2}+V_{i,5}\right)\right].
\end{align}
Since the nonlinear problem is effectively simplified to a two-mode system, we can rely on previous analysis to write the relevant polariton self-energies as follows:
\begin{align}
\Sigma_1^{R}(\omega) & = 4g_{211}^{2}  \frac{n_1-n_2}{\omega+\omega_{1}-\omega_{2}+i(\kappa_{1}+\kappa_{2})/2}, \label{SR1}\\
\Sigma_2^{R}(\omega) & =  4g_{211}^{2} \frac{n_1+1/2}{\omega-2\omega_{1}+i\kappa_{1}}, \label{SR2}
\end{align}
while $\Sigma_3^{R}(\omega) \simeq 0$.

%%%%%%%%%%%%%%%%%%%%%%%%%%%%%%%%%%%%%%%%%%%%%%%%%%%%%%%%%%%%%%%%%
\section{Resonant enhancement of nonlinear effects}\label{Sec: Resonant enhancement}
%%%%%%%%%%%%%%%%%%%%%%%%%%%%%%%%%%%%%%%%%%%%%%%%%%%%%%%%%%%%%%%%%

%%%%%%%%%%%%%%%%%%%%%%%%%%%%%%%%%%%%%%%%%%%%%%%%%%%%%%%%%%%%%%%%
\begin{figure}
\includegraphics[clip,width=0.48\textwidth]{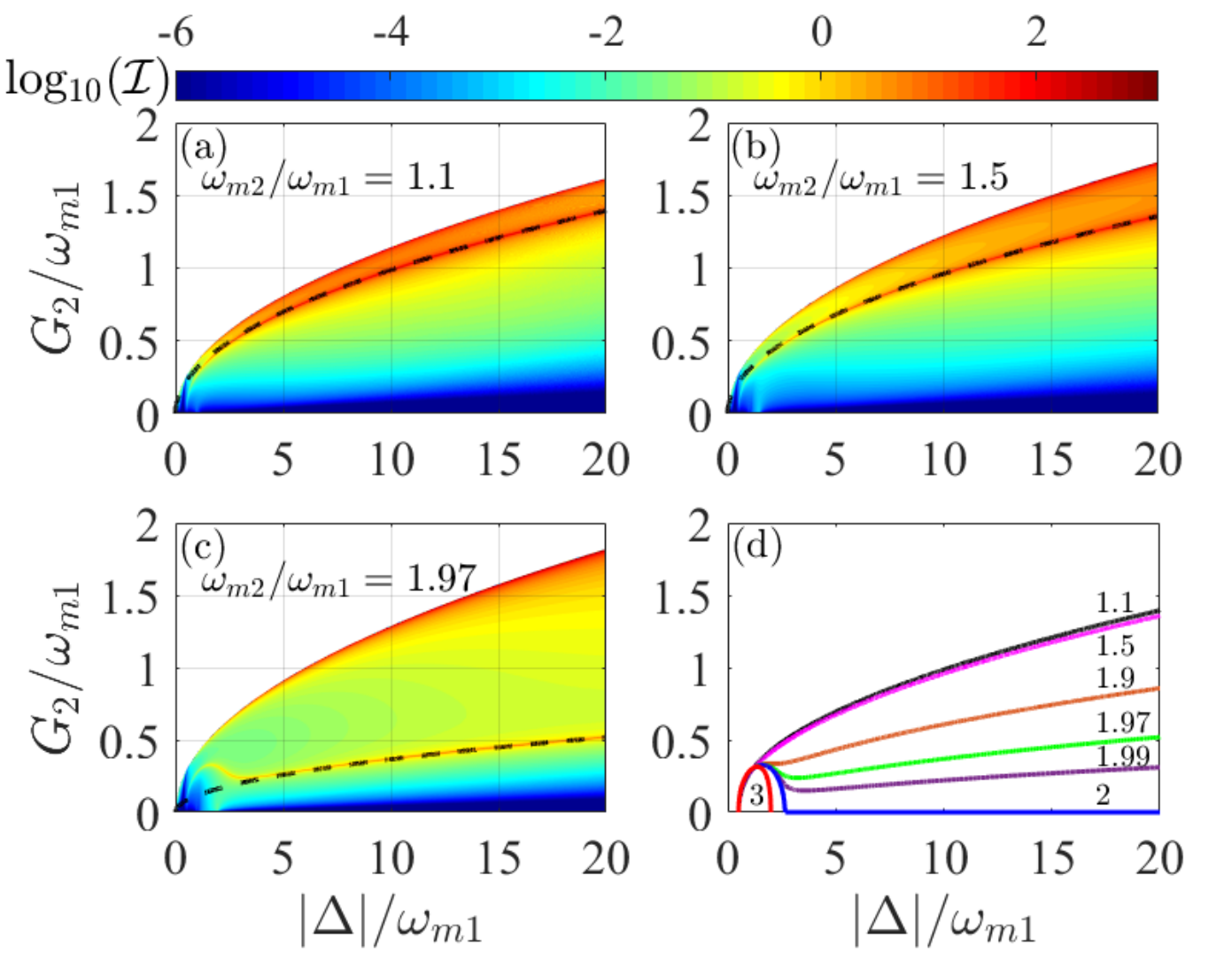}
\caption{Panels (a)-(c) show the dependence of $\mathcal{I}$ (quantifying nonlinear effects) on the dressed coupling $G_2=g_2 \sqrt{N}$ and detuning $|\Delta|$, at fixed values of $\omega_{m2}/\omega_{m1}$ (see figure) and  $g_1/g_2=1$ (implying $G_1=G_2$). The boundary of the colored regions is given by the stability condition Eq.~(\ref{eq:stability}). The black dashed curves are approximate resonant curves of the form $c\sqrt{\omega_{m1}|\Delta|}$, where the numerical prefactor $c$ is computed using the large-$|\Delta|$ expressions Eq.~(\ref{w12}). Panel (d) shows the resonant condition $\omega_2 = 2\omega_1$ obtained from the exact spectrum of Eq.~(\ref{M_matrix}). Each curve is computed at a different value of $\omega_{m2}/\omega_{m1}$, indicated in the plot. Other parameters (taking $\omega_{m1}=1$): $\kappa=0.02$, $\gamma_{1,2}=2 \times 10^{-6}$, $T=0$, and $g_{1}=g_{2}=2 \times 10^{-4}$.
\label{fig:Boundary_resonance}}
\end{figure}
%%%%%%%%%%%%%%%%%%%%%%%%%%%%%%%%%%%%%%%%%%%%%%%%%%%%%%%%%%%%%%%%

Examples quantifying nonlinear effects through the formalism described above are shown in Fig.~\ref{fig:Boundary_resonance}(a-c). There, we have evaluated $\rho_d$ using Eqs.~(\ref{eq:rho_d}), (\ref{GR}) and the perturbative self-energies Eqs.~(\ref{SR1}), (\ref{SR2}). More precisely, we plot (in logarithmic scale) the following quantity:
\begin{equation}\label{I_def}
\mathcal{I} = \max_{\omega}\left|\rho_{d}(\omega)-\rho_{d}^{0}(\omega)\right|,
\end{equation}
where $\rho_{d}^{0}(\omega)$ is the DOS without nonlinearity, i.e., assuming $H_{nl}=0$. $\mathcal{I}$ gives the maximum deviation of $\rho_{d}(\omega)$ induced by the nonlinear interactions over the whole spectrum. Considering $\mathcal{I}$ as function of the drive strength and detuning at fixed values of $\omega_{m1}$, $\omega_{m2}$, and  $g_{1,2}$, the most prominent feature within the stability region is the presence of a sharp line at which the nonlinear effects are enhanced. This condition corresponds to the resonance $\omega_2 = 2\omega_1$, which can be determined from the spectrum of Eq.~(\ref{M_matrix}). We show in Fig.~\ref{fig:Boundary_resonance}(d) how the resonant curves are modified by the ratio $\omega_{m2}/\omega_{m1}$. The resonant lines found from the spectrum match well the enhancement observed in panels (a-c).

Without any restrictions on the maximum photon number $N$, the most favorable regime is the one of large detuning $|\Delta| \gg \omega_{m1}$ and smaller ratios $\omega_{m2}/\omega_{m1} \gtrsim 1$. Similarly to the four-mode optomechanical ring discussed in Ref.~\onlinecite{Jin2018PRA}, a large value of $|\Delta|$  leads to enhanced nonlinear effects by inducing phonon-like polaritons with very small damping. Furthermore, as seen in Fig.~\ref{fig:Boundary_resonance}, a smaller value of $\omega_{m2}/\omega_{m1} $ allows the resonant curve to approach the boundary of the unstable regime. We note, however, that the optimization of $\omega_{m2}/\omega_{m1}$ is nontrivial since for $\omega_{m2}/\omega_{m1}=1$ nonlinear effects drop to small values (see Appendix~\ref{appendix_equal_wm}).

In the spectral domain, the DOS is characterized by three nearly Lorenzian polariton peaks. As it turns out, the largest changes in the DOS occur at the two lower peaks, with frequencies $\omega_{1,2}$. The high-frequency polariton is not involved in the resonant scattering process and the corresponding peak at $\omega_3$ is hardly affected by nonlinear effects. To quantify the change of DOS at $\omega_{1,2}$ we can evaluate Eq.~(\ref{eq:rho_d}) at the relevant polariton frequencies and keep only the dominant contribution $\propto G^{R}[c_j,c_j^{\dagger};\omega_j]$, thus obtaining the approximate expressions:
\begin{equation} \label{rho_d_approx_Ceff}
\rho_{d}(\omega_{1})\approx \frac{2V_{31}^2}{\pi\kappa_{1}}\frac{1}{1+C_{\mathrm{eff},1}}, \quad \rho_{d}(\omega_{2})\approx \frac{2V_{32}^2}{\pi\kappa_{2}}\frac{1}{1+C_{\mathrm{eff,2}}},
\end{equation}
where the effective cooperativities appearing in the denominators are given by
\begin{equation} \label{eq:C_eff}
C_{\mathrm{eff,1}}=\frac{16g_{211}^{2}(n_{1}-n_{2})}{\kappa_{1}(\kappa_{1}+\kappa_{2})}, \qquad C_{\mathrm{eff,2}}=\frac{4g_{211}^{2}(1+2n_1)}{\kappa_{1}\kappa_{2}}.
\end{equation}
The $C_{\mathrm{eff},j}$ are directly related to the relative changes of DOS induced by the nonlinear interaction, since $\left|\rho_{d}(\omega)-\rho_{d}^{0}(\omega)\right|/\rho_d(\omega_j) \simeq C_{\mathrm{eff},j}$ (usually, $C_{\mathrm{eff},j}\ll 1$, due to the weakness of nonlinear interactions). In the following we will mostly discuss $C_{\mathrm{eff,2}}$, which is more directly comparable to the two-mode system (in the two-mode system, $n_{i} \ll 1$ and the nonlinear effects at the lower polariton are small). However, we will discuss at the end of Sec.~\ref{Sec:Result_2} that for the three-mode system the two cooperatives are similar at the optimal point, thus the behavior of $C_{\mathrm{eff,2}}$ is also representative for  $C_{\mathrm{eff,1}}$ (see also Fig.~\ref{fig:DOS}).

%%%%%%%%%%%%%%%%%%%%%%%%%%%%%%%%%%%%%%%%%%%%%%%%%%%%%%%%%%
\begin{figure}
\includegraphics[clip,width=0.48\textwidth]{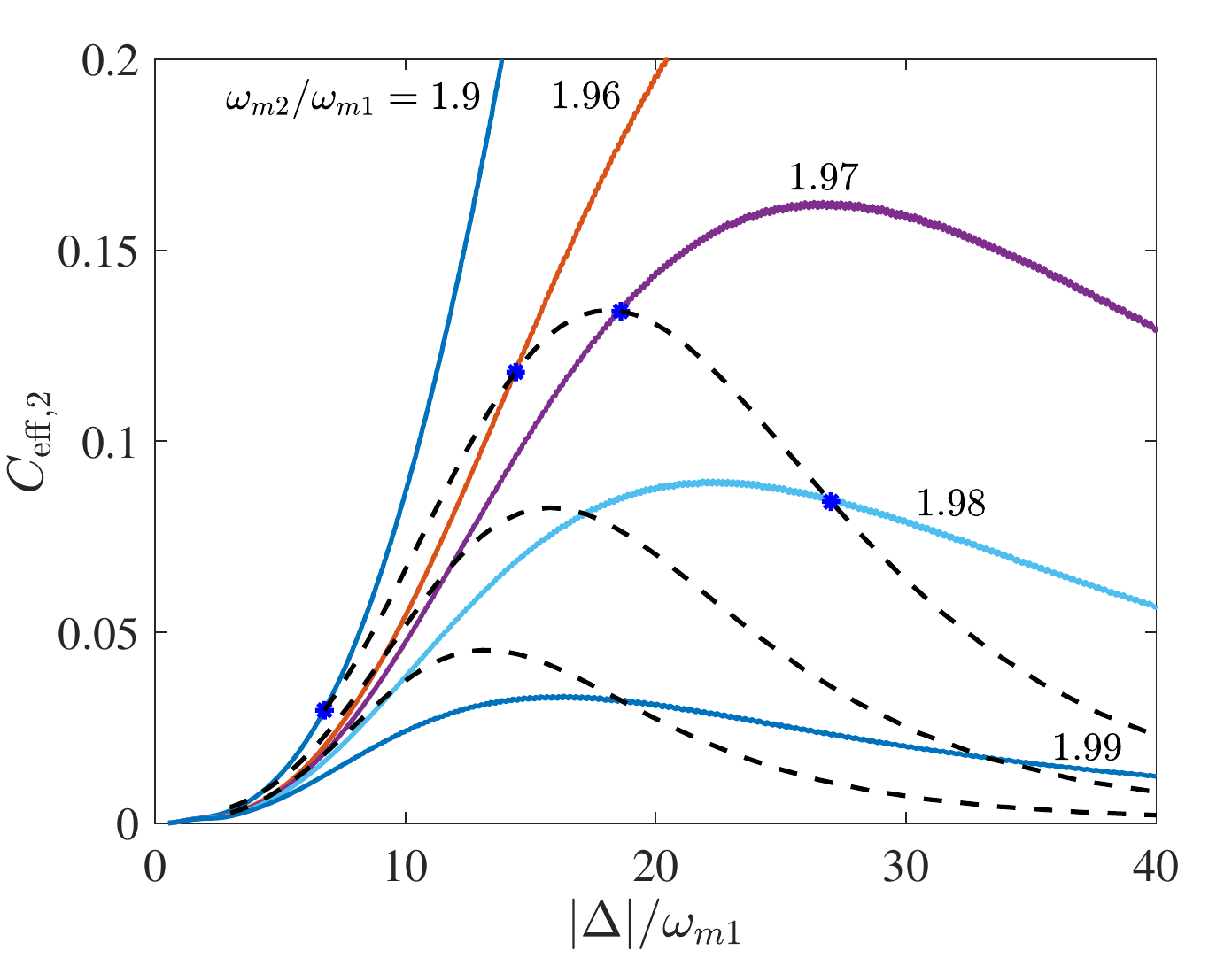}
\caption{Dependence of $C_{\mathrm{eff,2}}$ on $|\Delta|$ along the resonant curves, at different values of the ratio $\omega_{m2}/\omega_{m1}$ (as indicated for each solid curve). The dashed lines correspond to the condition $G_{1,2}\leq G_{\rm max}$, with $G_{\rm max}/\omega_{m1} = 0.5, 0.4, 0.3 $ (top to bottom). For each solid curve, the blue star marks the maximum value of $|\Delta|/\omega_{m1}$ allowed by $G_{1,2} \leq 0.5 \omega_{m1}$. Other parameters are the same of Fig.~\ref{fig:Boundary_resonance}.}
\label{fig:C_eff_resonance_total}
\end{figure}
%%%%%%%%%%%%%%%%%%%%%%%%%%%%%%%%%%%%%%%%%%%%%%%%%%%%%%%%%%

Examples of the numerically evaluated $C_{\mathrm{eff,2}}$ as function of $|\Delta|$ along the resonant curves are shown in Fig.~\ref{fig:C_eff_resonance_total}. The dependence of $C_{\mathrm{eff,2}}$ is non-monotonic, similarly to the four-mode chain \cite{Jin2018PRA}. The decrease of  $C_{\mathrm{eff,2}}$ at large $|\Delta|$ is due to the saturation of the polariton damping $\kappa_{1,2}$ to the bare mechanical dissipation rates, $\gamma_{1,2}$ (a more detailed discussion is provided later on). We also see that, in agreement with the previous discussion, the curves with a smaller $\omega_{m2}/\omega_{m1}$ lead to larger values of $C_{\mathrm{eff,2}}$. However, this advantageous behavior does not take into account any practical limitation on the maximum achievable dressed coupling strength.

Since it is  difficult to realize values of $G_{1,2}$ approaching the mechanical frequency (the regime of ultrastrong coupling \cite{Peterson2019PRL}), we also consider the effect of a upper cutoff $G_{1,2}\leq G_{\rm max}$, where $G_{\rm max}$ is smaller than $\omega_{m1}$. This restriction is illustrated by the dashed curves of Fig.~\ref{fig:C_eff_resonance_total}. As seen, a system with smaller value of $\omega_{m2}/\omega_{m1}$ suffers a strong reduction on the maximum value of $|\Delta|$, marked by a blue star for each solid curve (considering $G_{\rm max}=0.5 \omega_{m1}$). Therefore, one has to strike a compromise between the generally advantageous effect of reducing $\omega_{m2}/\omega_{m1}$ and the more restrictive range of $|\Delta|$. The optimal choice of $\omega_{m2}/\omega_{m1}$ is generally slightly below $2$. For example, we see that the purple curve with $\omega_{m2}/\omega_{m1} = 1.97$ hits the upper dashed boundary close to its maximum, thus represents the optimal choice when $G_{\rm max}=0.5 \omega_{m1}$.

%%%%%%%%%%%%%%%%%%%%%%%%%%%%%%%%%%%%%%%%%%%%%%%%%%%%%%%%%%
\begin{figure}
\includegraphics[clip,width=0.48\textwidth]{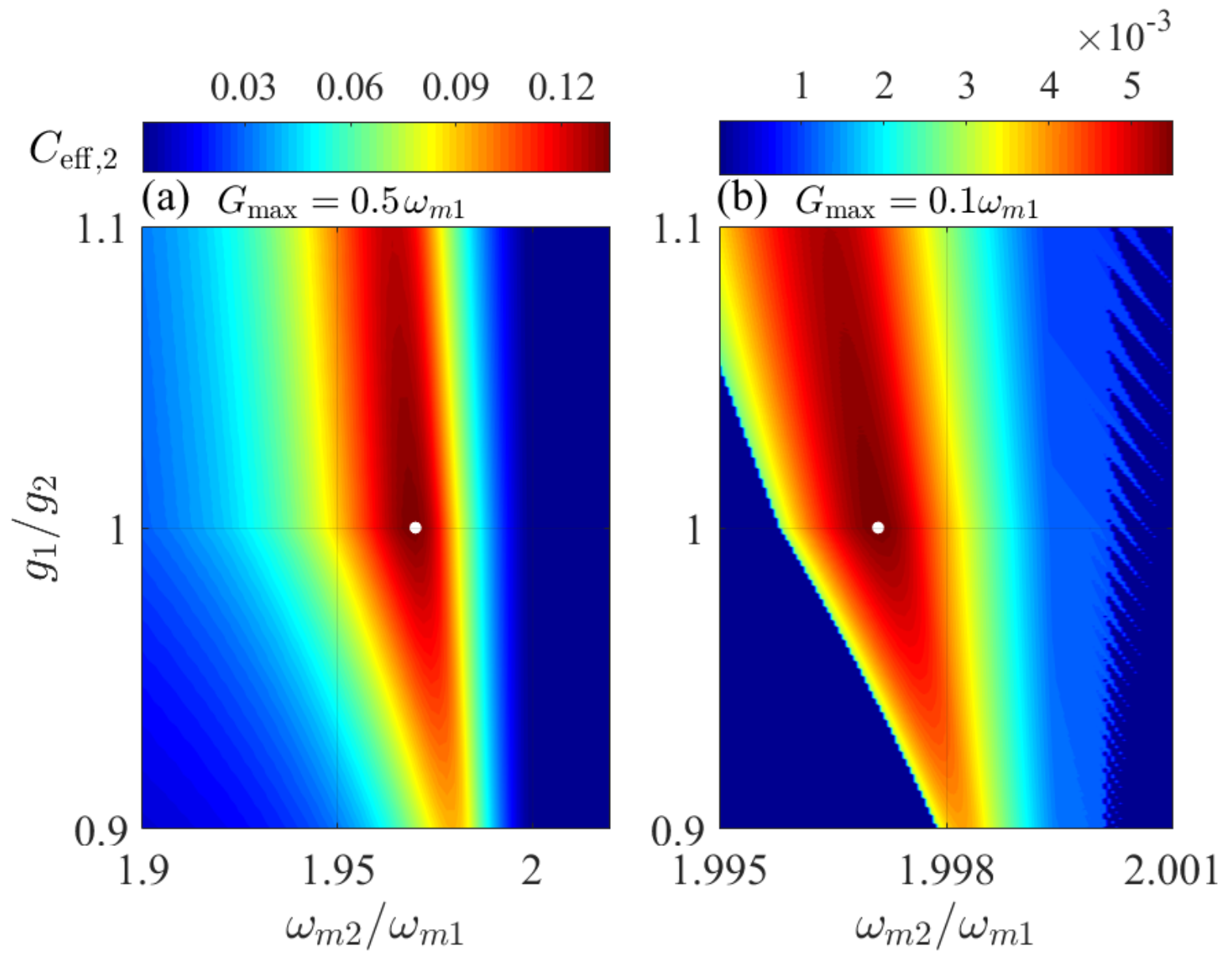}
\caption{Maximum value of $C_{\mathrm{eff,2}}$ (i.e., optimized over $\Delta$) for different ratios of $\omega_{m2}/\omega_{m1}$ and $g_1/g_2$, with the restriction $\textrm{max}[G_1, G_2]\leq G_{\rm max}$. The dots mark the optimal values $\tilde{C}_{\mathrm{eff},2}$. We used the following parameters (in units of $\omega_{m1}$): $\kappa=0.02$, $\textrm{max}[g_1, g_2] =2\times 10^{-4}$, and $\gamma_{1,2}=2\times 10^{-6}$.
\label{fig:Fig_4_C_eff_Vs_g1_omega_m2}}
\end{figure}
%%%%%%%%%%%%%%%%%%%%%%%%%%%%%%%%%%%%%%%%%%%%%%%%%%%%%%%%%%

We also show in Fig.~\ref{fig:Fig_4_C_eff_Vs_g1_omega_m2} a density plot of the maximum $C_{\mathrm{eff,2}}$ (i.e., after optimizing over $\Delta$) as function of $g_1/g_2$ and  $\omega_{m2}/\omega_{m1} $, for two different choices of $ G_{\rm max}$. The largest value is marked by a white dot and satisfies $g_1=g_2$. Instead, the optimal value of $\omega_{m2}/\omega_{m1} $ is always slightly below 2 but depends on $ G_{\rm max}$.

Finally, we note that in Fig.~\ref{fig:C_eff_resonance_total} the largest value of $C_{\rm eff,2}$ (the maxima of the dashed curves) are attained in the regime of large detunings $|\Delta|\gg \omega_{m1}$. In the next Sec.~\ref{Sec:Result_2} we explore more explicitly this limit, which allows us to obtain analytical expressions through a perturbative treatment and identify the most relevant parametric dependences.

%%%%%%%%%%%%%%%%%%%%%%%%%%%%%%%%%%%%%%%%%%%%%%%%%%%%%%%%%%
\section{Large detuning limit}\label{Sec:Result_2}
%%%%%%%%%%%%%%%%%%%%%%%%%%%%%%%%%%%%%%%%%%%%%%%%%%%%%%%%%%

By assuming $|\Delta| \gg \omega_{m1}, \omega_{m2}$ and relatively small dressed couplings, $G_{1,2}\lesssim \omega_{m1}$, we can diagonalize Eq.~(\ref{M_matrix}) perturbatively and simplify the expressions of the linear transformation matrices $U$ and $V$ (see Appendix~\ref{Appendix_V} for details). Under these conditions, the two lower polariton modes are  phonon-like, i.e., they are weakly mixed with the optical cavity. Their dampings take the approximate form:
\begin{equation}
\kappa_{i} \simeq \frac{4G_{i}^{2}\omega_{mi}}{\left|\Delta\right|^{3}} \kappa+\gamma_{i},
\label{pert_theory_kappa}
\end{equation}
where $i=1,2$. The total decay rate is the sum of an optical contribution, due to the small mixing to the optical cavity, and the regular mechanical damping. With the same approach we also obtain the relevant occupation number and effective interaction as follows:
\begin{equation}
n_{1}    \simeq \left( \frac{\gamma_1 \Delta^2}{\kappa G_1^2} +4 \frac{\omega_{m1}}{|\Delta|}\right)^{-1} ,
~~ g_{211}  \simeq 3g_1 \frac{G_{1}G_{2}}{\Delta^{2}}.
\label{pert_theory_n_g}
\end{equation}
These expressions can be substituted in Eq.~(\ref{eq:C_eff}), giving:
\begin{equation}
C_{\mathrm{eff,2}}\simeq\frac{72 \kappa g^2 G_2^6 |\Delta|^3 \left(1+\frac{\gamma_1 \Delta^2}{2\kappa G_2^2} +2 \frac{\omega_{m1}}{|\Delta|} \right) }
{\left( 4G_2^2\omega_{m1}\kappa +\gamma_1 |\Delta|^3\right)^2\left( 8G_2^2\omega_{m1}\kappa +\gamma_2 |\Delta|^3\right)},
\label{eq:C_eff_2}
\end{equation}
where we used that, as in Fig.~\ref{fig:Fig_4_C_eff_Vs_g1_omega_m2}, the optimal point occurs for $g_1=g_2 \equiv g$ (thus $G_1=G_2$) and $\omega_{m2} \simeq 2 \omega_{m1}$. Note that the expression of $C_{\mathrm{eff,2}}$ is derived at resonance, implying that $G_2$ and $\Delta$ in Eq.~(\ref{eq:C_eff_2}) must satisfy the resonant condition. From the expressions of $\omega_{1,2}$ in Eq.~(\ref{w12}), and taking into account $g_2/g_1=1$ and $\omega_{m2}/\omega_{m1}\simeq 2$, we obtain the following approximate relationship:
\begin{equation}\label{eq:Resonance condition_text}
G_2 \simeq \sqrt{\left(\omega_{m1}- \frac{\omega_{m2}}{2}\right) |\Delta|}.
\end{equation}

We now proceed to optimize Eq.~(\ref{eq:C_eff_2}) and set $G_2=G_{\rm max}$. For simplicity we consider equal mechanical dampings $\gamma_{1,2}=\gamma$ (extension to unequal dampings is straightforward) and work under the assumption that $1+\gamma_1 \Delta^2/(2\kappa G_1^2) +2 \omega_{m1}/|\Delta| \simeq 1$ in the numerator of Eq.~(\ref{eq:C_eff_2}). The latter approximation implies:
\begin{equation}
R  = \left(\frac{G_{\mathrm{max}}}{\omega_{m1}}\right)^{2} \frac{\kappa}{\gamma} \gg 1.
\label{eq:C_eff_max_1}
\end{equation}
As it will become clear in the following, $R$ is an important parameter controlling the enhancement of $C_{\mathrm{eff,2}}$ with respect to a two-mode system. Therefore,  having a large value of $R$ is desirable. Under these assumptions, Eq.~(\ref{eq:C_eff_2}) simplifies to:
\begin{equation}
C_{\mathrm{eff,2}}\simeq\frac{72 \kappa g^2 G_{\rm max}^6 |\Delta|^3 }
{\left( 4G_{\rm max}^2\omega_{m1}\kappa +\gamma |\Delta|^3\right)^2\left( 8G_{\rm max}^2\omega_{m1}\kappa +\gamma |\Delta|^3\right)},
\label{eq:C_eff_2_simplified}
\end{equation}
where the resonant condition reads:
\begin{equation} \label{Delta_resonant}
|\Delta| \simeq \frac{G_{\rm max}^2}{\omega_{m1}}\left(1-\frac{\omega_{m2}}{2\omega_{m1}} \right)^{-1}.
\end{equation}

%%%%%%%%%%%%%%%%%%%%%%%%%%%%%%%%%%%%%%%%%%%%%
\begin{figure}
\includegraphics[clip,width=0.35\textwidth]{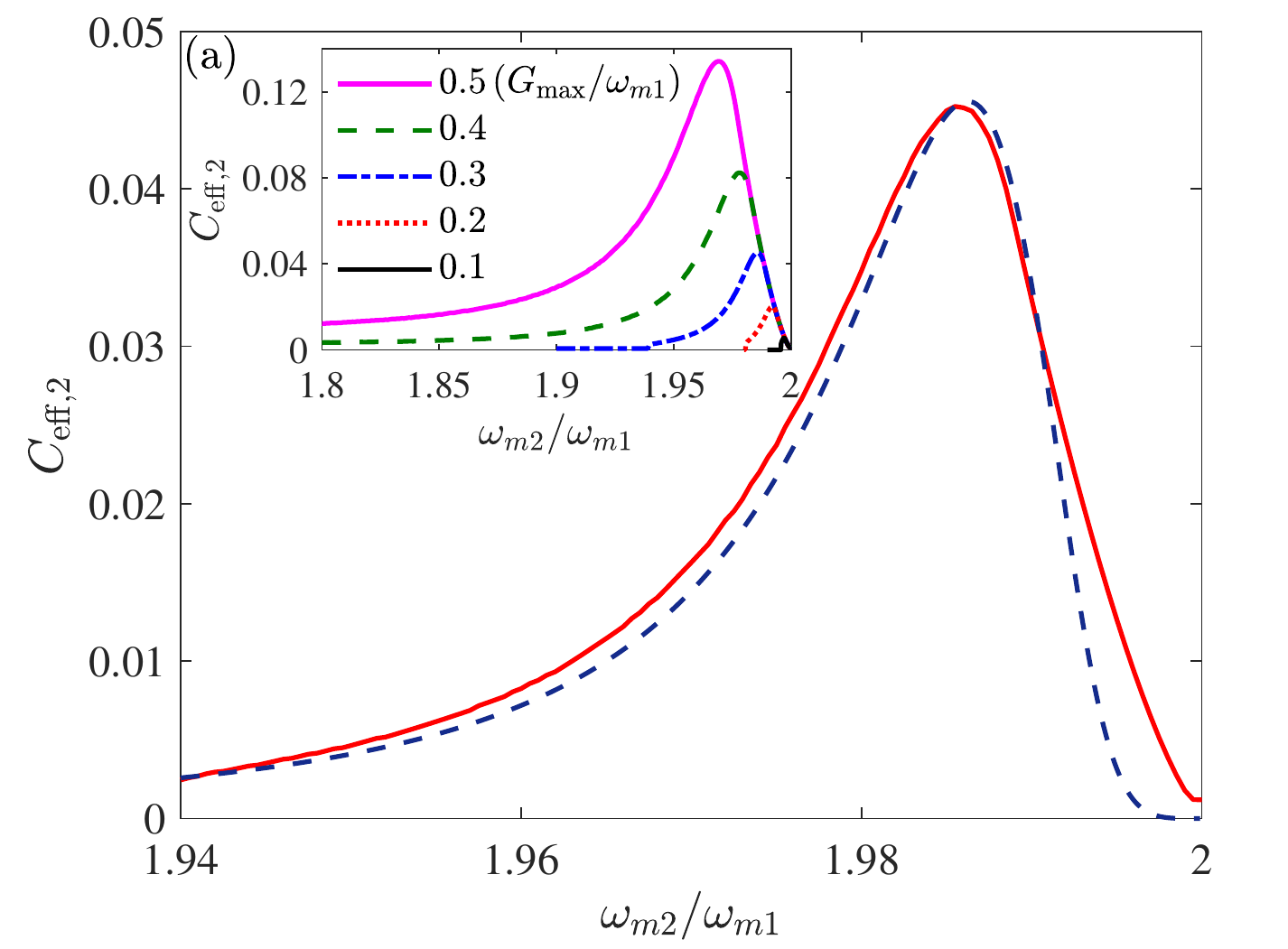}
\includegraphics[clip,width=0.35\textwidth]{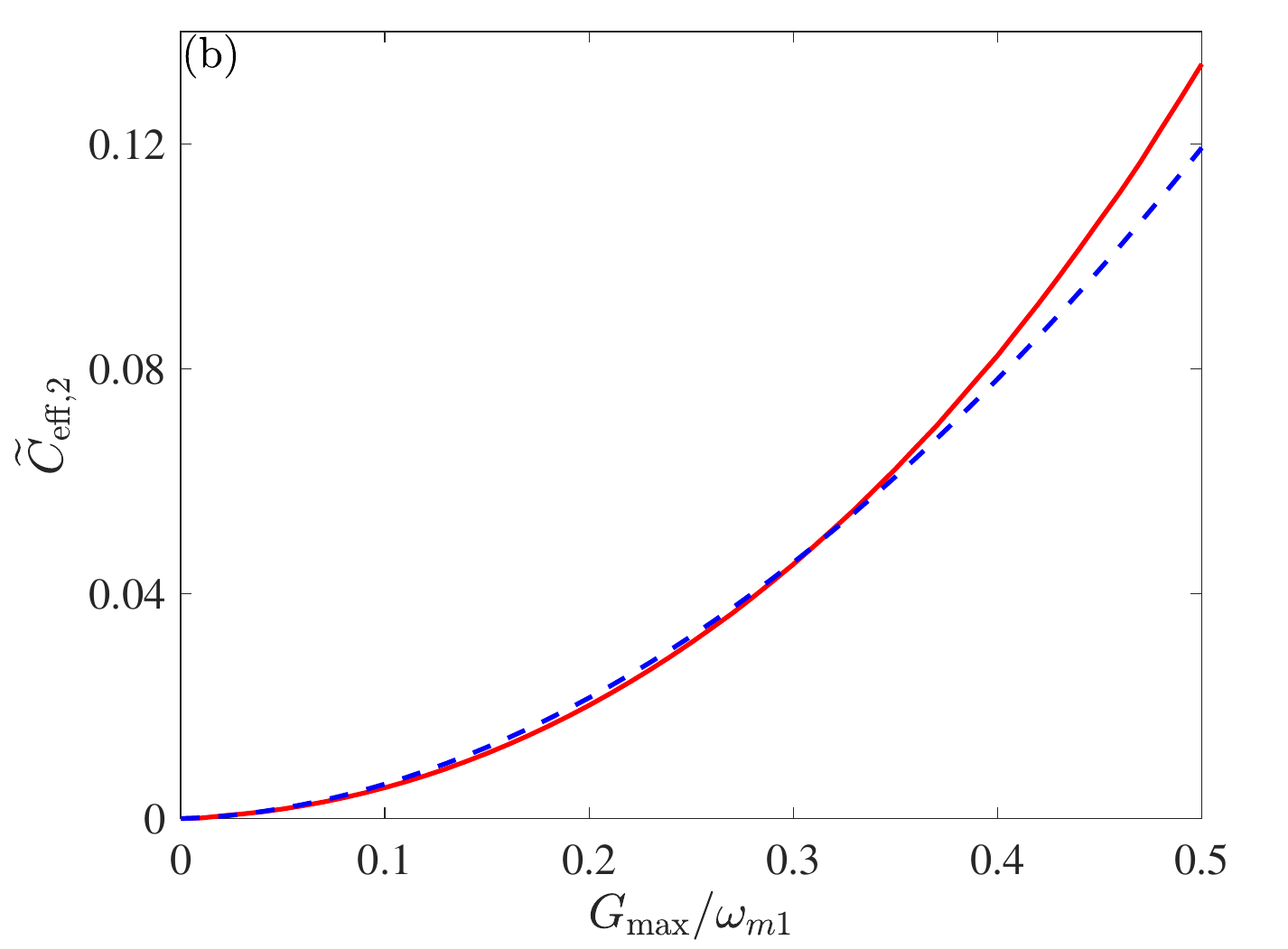}
\caption{ (a): Dependence of $C_{\mathrm{eff,2}}$ on $\omega_{m2}/\omega_{m1}$, obtained after optimizing $\Delta$ with $g_1/g_2=1$ and $G_{1,2}\leq G_{\rm max}$. The main panel is for $G_{\rm max}/\omega_{m1}=0.3$ and presents a comparison of the numerical curve (solid) to the approximate treatment (dashed), described by Eqs.~(\ref{eq:C_eff_2}) and (\ref{Delta_resonant}). The curves in the inset are obtained numerically and illustrate the effect of changing $G_{\rm max}/\omega_{m1}$. In particular, the $G_{\rm max}/\omega_{m1}=0.1$ and 0.5 curves are line cuts of Fig.~\ref{fig:Fig_4_C_eff_Vs_g1_omega_m2}. (b): Dependence of the fully optimized cooperativity  $\widetilde{C}_{\mathrm{\mathrm{eff,2}}} $ on $G_{\rm max}/\omega_{m1}$. The dashed curve is the approximate Eq.~(\ref{eq:C_eff_optimal}). In both panels we used (in units of $\omega_{m1}$): $\kappa=0.02$, $\gamma_{1,2}=2\times 10^{-6}$, and $g_{1,2}=2\times 10^{-4}$.
\label{fig:The-comparison}}
\end{figure}
%%%%%%%%%%%%%%%%%%%%%%%%%%%%%%%%%%%%%%%%%%%%%

Equations~(\ref{eq:C_eff_2_simplified}) and (\ref{Delta_resonant}) give the maximum $C_{\mathrm{eff,2}}$ at given mechanical frequencies. A comparison of the approximate result and the numerical evaluation is provided in Fig.~\ref{fig:The-comparison}(a), showing good agreement. As noted previously in Fig.~\ref{fig:Fig_4_C_eff_Vs_g1_omega_m2}, there is an optimal value of the ratio $\omega_{m2}/\omega_{m1}$, giving the largest nonlinear effect. To find this value we further optimize Eq.~(\ref{eq:C_eff_2_simplified}) with respect to $\omega_{m2}$ and obtain:
\begin{equation} \label{eq:C_eff_optimal}
\widetilde{C}_{\mathrm{\mathrm{eff,2}}}  \simeq  (c_1R+c_2R^{2/3} + \ldots) \left(\frac{g}{\kappa}\right)^{2} ,
\end{equation}
where the coefficients are $c_1 = 9(5\sqrt{5}-11)/4  \approx 0.41$ and $c_2 = 9(7-3\sqrt{5})/\sqrt[3]{16(\sqrt5-1)} \approx 0.97$. The numerical prefactor in Eq.~(\ref{eq:C_eff_optimal}) is expressed through powers of the large enhancement factor $R$ and the fully optimized cooperativity $\widetilde{C}_{\mathrm{\mathrm{eff,2}}} $ exhibits a monotonic dependence on $G_{\rm max}$, due to the quadratic increase of $R$ with $G_{\rm max}$ [see Eq.~(\ref{eq:C_eff_max_1})]. In Fig.~\ref{fig:The-comparison}(b) we show that the increase of $\widetilde{C}_{\mathrm{eff,2}}$ is well described by the approximate Eq.~(\ref{eq:C_eff_optimal}).

Finally, we find that the maximum of $\widetilde{C}_{\mathrm{\mathrm{eff,2}}}$ occurs at
\begin{equation}\label{Delta_max_precise}
\frac{\omega_{m2}}{\omega_{m1}} \simeq 2- c_0 \left(\frac{\gamma}{\kappa}\right)^{1/3}\left(\frac{G_{\rm max}}{\omega_{m1}}\right)^{4/3},
\end{equation}
where $c_0= (\sqrt{5}+1)^{1/3} \simeq 1.5$. Since usually $\gamma \ll \kappa$, the second term of Eq.~(\ref{Delta_max_precise}) represents a small deviation from  $\omega_{m2} = 2\omega_{m1}$. The inset of Fig.~\ref{fig:The-comparison}(a) shows that a  larger dressed optomechanical coupling $G_{\rm max}$ causes the optimal ratio $\omega_{m2}/\omega_{m1}$ to move farther away from $\omega_{m2}=2\omega_{m1}$, besides allowing for more prominent nonlinear effects, which is in agreement with Eq.~(\ref{Delta_max_precise}).

\subsection{Comparison to a two-mode system}\label{Appendix_parameters}

We perform now a more specific comparison to the two-mode setup, where the optimal point is at $\Delta \simeq -2\omega_{m} $, leading to ~\cite{Jin2018PRA}:
\begin{equation}
C_{\rm eff}  \simeq \frac{45}{8} \frac{g^2}{\kappa^2}. ~~~{\rm (two~modes)}
\end{equation}
As a reference we consider parameters from a very recent electromechanical setup using a 3D superconducting cavity, which allowed achieving ultrastrong parametric couplings of order $G_{\rm max} \sim 0.4 \omega_m$ \cite{Peterson2019PRL}. With the parameters listed in Table~\ref{table_parameters}, we estimate that the ultrastrong-coupling regime would allow for a potentially large enhancement factor of order $R \simeq 6\times 10^3$. In fact, from Eq.~(\ref{eq:C_eff_optimal}), the figure of merit for the nonlinear effects would be a more accessible $\widetilde{C}_{{\rm eff},2} \simeq  0.5 \times 10^{-4}$ in the three-mode system, instead of $C_{\rm eff} \simeq 10^{-7}$ for the two-mode setup.

%%%%%%%%%%%%%%%%%%%%%%%%%%%%%%%%%%%%%%%%%%%%%
\begin{table}
\centering
\caption{Parameters from two specific setups.}
\label{table_parameters}
\begin{tabularx}{0.49\textwidth}{XXX}
\hline
\hline
                    &  Peterson \emph{et al.}~\cite{Peterson2019PRL}  & Teufel \emph{et al.}~\cite{Teufel2011Nature} \\
\hline
$\omega_{c}/2\pi$    & 6.506 GHz & 7.47  GHz  \\
$\kappa/2\pi$      & 1.2 MHz  & 170 kHz \\
$\omega_{m}/2\pi$   & 9.696 MHz & 10.69 MHz \\
$\gamma/2\pi$      & $31\pm 1$ Hz  & 30 Hz  \\
$g/2\pi$            & $167\pm 2$ Hz & 230 Hz \\
$G_{\rm max}/2\pi $  & 3.83 MHz  & 0.5 MHz  \\
\hline
\hline
\end{tabularx}
\end{table}
%%%%%%%%%%%%%%%%%%%%%%%%%%%%%%%%%%%%%%%%%%%%%

It is also instructive to consider parameters from an electromechanical setup with
lumped elements (second column of Table~\ref{table_parameters}) and a much smaller $G_{\rm max} \sim  0.05 \omega_m $. Here, also due to the
weaker cavity damping, we only have $R \simeq 12$. Indeed, we
estimate that the three-mode system could achieve $\widetilde{C}_{{\rm eff},2}
 \simeq  2 \times 10^{-5}$, similar to $C_{\rm eff} \simeq 10^{-5}$ of the
two-mode setup.

The final values for $\widetilde{C}_{{\rm eff},2} $ in the two scenarios are similar, despite the great difference in $(g/\kappa)^2$. In the first example, the value of $\widetilde{C}_{{\rm eff},2} $  suffers from the relatively large damping of the microwave cavity. Improving that parameter to the $\sim 100$~kHz range would result in a  much larger value of $(g/\kappa)^2$, thus approaching $\widetilde{C}_{{\rm eff},2} \simeq  10^{-3}$ in the three-mode system. We also note that the working point of Ref.~\cite{Peterson2019PRL} is at $\Delta = -\omega_m$, when the dressed optomechanical coupling is limited by the optomechanical instability to $G_{\rm max}  < 0.5 \omega_m$. However, here we consider large values of $|\Delta|$ and the onset of the instability is less restrictive on $G_{\rm max}$. This is potentially beneficial to the enhancement of nonlinear effects, due to the strong dependence of $R\propto (G_{\rm max}/\omega_m)^2$.

%%%%%%%%%%%%%%%%%%%%%%%%%%%%%%%%%%%%%%%%%%%%%
%\begin{figure}
%\includegraphics[clip,width=0.35\textwidth]{C_eff_compare_Nature}
%\includegraphics[clip,width=0.35\textwidth]{C_eff_compare_PRL}
%\caption{ Comparison between the two-mode system and our three-mode system. The red solid curve show the dependence of $C_{\mathrm{eff,2}}$ on $\omega_{m2}/\omega_{m1}$, obtained after optimizing $\Delta$ with $g_1/g_2=1$ and $G_{1,2}\leq G_{\rm max}$. The top horizontal black dot-dashed line and the bottom blue dashed line mark the approximately strongest nonlinear effects with optimized parameters for three-mode (Eq.~(\ref{eq:C_eff_optimal})) and two-mode systems, respectively. The parameters in panels (a) and (b) are extracted from Refs.~\cite{Teufel2011Nature} and ~\cite{Peterson2019PRL}, respectively, i.e., the experimental parameters are given in Table~\ref{parameters}.
%\label{Three_Two_compare}}
%\end{figure}
%%%%%%%%%%%%%%%%%%%%%%%%%%%%%%%%%%%%%%%%%%%%%
%The top horizontal black dot-dashed line is the approximately strongest nonlinear effects Eq.~(\ref{eq:C_eff_optimal}).
%The bottom blue dashed line marks the approximately maximum cooperativity of two-mode system, see Eq.~(\ref{two-mode}).

\subsection{Physical origin of the enhancement}

We have seen that in a two-mode optomechanical system $C_{\rm eff} \sim (g/\kappa)^{2}$ \cite{Lemonde2013PRL, Lemonde2015PRA, Borkje2013PRL}. Therefore, in Eq.~(\ref{eq:C_eff_optimal}) we can identify $c_1 R$ as the approximate enhancement factor induced by the three-mode setup. As shown in the previous section, a large value of $R$ can be realized for relatively small values of $G_{\rm max}$, due to the typical smallness of $\gamma/\kappa$.

To understand the physical origin of the enhancement, it is useful to first examine the non-monotonic dependence of $C_{\mathrm{eff,2}}$ as function of $\Delta$, shown in Fig.~\ref{fig:C_eff_resonance_total}. As it turns out, the maximum of $C_{\mathrm{eff,2}}$ occurs approximately when the mechanical damping $\gamma$ becomes comparable to the induced optical damping $\kappa_i^{\rm opt}$ [i.e., the first term of Eq.~(\ref{pert_theory_kappa})]. By taking into account the resonant condition, one has:
\begin{equation}\label{kappa_reduced_approx}
\kappa_1^{\rm opt}\simeq \left(1-\frac{\omega_{m2}}{2\omega_{m1}} \right)\frac{4 \omega_{m1}^2}{\Delta^2}\kappa,
\end{equation}
while $\kappa_2^{\rm opt}\simeq 2 \kappa_1^{\rm opt}$. Then, imposing $\kappa_i^{\rm opt} \sim \gamma$ we estimate that the maximum of $C_{\mathrm{eff,2}}$ occurs at:
\begin{equation}\label{Delta_maximum}
|\Delta^*|\sim \sqrt{\frac{\kappa}{\gamma}\left(1-\frac{\omega_{m2}}{2\omega_{m1}} \right)}\omega_{m1}.
\end{equation}

To see that this value of $\Delta$ is reasonable we first assume $|\Delta|\ll |\Delta^*|$, when the mechanical damping can be neglected. Setting $\gamma_{1,2}=0$ in Eq.~(\ref{eq:C_eff_2}) and also neglecting $2\omega_{m1}/|\Delta|$ in the numerator (since $|\Delta| \gg \omega_{m1}$), we obtain a monotonically increasing function:
\begin{equation}\label{eq:C_eff_2_small_Delta}
C_{\mathrm{eff,2}}\simeq\frac{9}{16} \left(\frac{g}{\kappa}\right)^2 \left(\frac{|\Delta|}{\omega_{m1} } \right)^3 .
\end{equation}
The cubic dependence of $C_{\mathrm{eff,2}}$ on $\Delta$ has the following origin: In Eq.~(\ref{eq:C_eff}), the product $\kappa_1 \kappa_2$ of damping rates contribute to an enhancement factor $\sim (|\Delta|/\omega_{m1})^4$ to the effective cooperativity. The increase of $n_1$ is $\sim |\Delta|/(4\omega_{m1})$, and can also be attributed to the improved coherence of the polariton modes [i.e., to the small $\kappa_i$ denominator in Eq.~(\ref{n_i})]. On the other hand, as expected, the effective interaction $g_{211}$ is suppressed as $|\Delta|^{-1}$. The final result is the cubic enhancement factor $(|\Delta|/\omega_{m1})^3$.

The initial growth of $C_{\mathrm{eff,2}}$ with $|\Delta|$ is due to the reduction of the hybridization with the optical modes, which improves the coherence properties of the $i=1,2$ polaritons. This mechanism clearly breaks down when $|\Delta|\gg |\Delta^*|$ and $\kappa_i \simeq \gamma_i$ is a constant. In that regime, the behavior of $C_{\mathrm{eff,2}}$ is dominated by the decreasing interaction strength $g_{211}$ between almost purely phononic polaritons, resulting in the non-monotonic dependence.

A rough estimate of the maximum value of $C_{\mathrm{eff,2}}$ is obtained by evaluating Eq.~(\ref{eq:C_eff_2_small_Delta}) at $\Delta=\Delta^*$:
\begin{equation}\label{C_eff_max_estimate}
C_{\mathrm{eff,2}} \lesssim \left(\frac{g}{\kappa}\right)^2 \left[\frac{\kappa}{\gamma}\left(1-\frac{\omega_{m2}}{2\omega_{m1}}\right) \right]^{3/2} .
\end{equation}
One can see from the presence of the factor $(1-\frac{\omega_{m2}}{2\omega_{m1}})^{3/2}$ that, in principle, it is advantageous to decrease the ratio $\omega_{m2}/\omega_{m1}$ away from $\omega_{m2} = 2 \omega_{m1}$. However, as discussed, one should take into account practical limitations on the achievable $G_{1,2}$.

With $G_{1,2}\leq G_{\rm max}$, the maximum allowed value of $|\Delta|$ is given by Eq.~(\ref{Delta_resonant}), where the factor $1-\frac{\omega_{m2}}{2\omega_{m1}}$ appears in the denominator (i.e., the allowed range shrinks by reducing $\omega_{m2}/\omega_{m1}$). An approximate criterion to estimate the optimal $\omega_{m2}/\omega_{m1}$ is to impose that the range of allowed values of $\Delta$ extends roughly up to the maximum in $C_{\mathrm{eff,2}}$. Equating Eqs.~(\ref{Delta_resonant}) and (\ref{Delta_maximum}) yields:
\begin{equation}\label{estimate_w12_ratio}
1-\frac{\omega_{m2}}{2\omega_{m1}} \sim \left(\frac{\gamma}{\kappa}\right)^{1/3}\left(\frac{G_{\rm max}}{\omega_{m1}}\right)^{4/3},
\end{equation}
and substituting this estimate in Eq.~(\ref{C_eff_max_estimate}):
\begin{equation}\label{eq:C_eff_max_estimate}
\widetilde{C}_{\mathrm{eff,2}}\sim \left(\frac{g}{\kappa}\right)^2 \frac{\kappa}{\gamma}\left(\frac{G_{\rm max}}{\omega_{m1}}\right)^2 .
\end{equation}
The above Eqs.~(\ref{estimate_w12_ratio}) and (\ref{eq:C_eff_max_estimate}) are in agreement with the more precise Eqs.~(\ref{Delta_max_precise}) and~(\ref{eq:C_eff_optimal}), respectively. Through this discussion, we see that the optimal values arise from a competition between the reduction in the effective optical damping at large $|\Delta|$ and the presence of a residual mechanical damping, together with practical restrictions in achieving sufficiently large dressed optomechanical couplings.

%%%%%%%%%%%%%%%%%%%%%%%%%%%%%%%%%%%%%%%%%%%%%%%%%%%%%%
\begin{figure}
\includegraphics[clip,width=9cm]{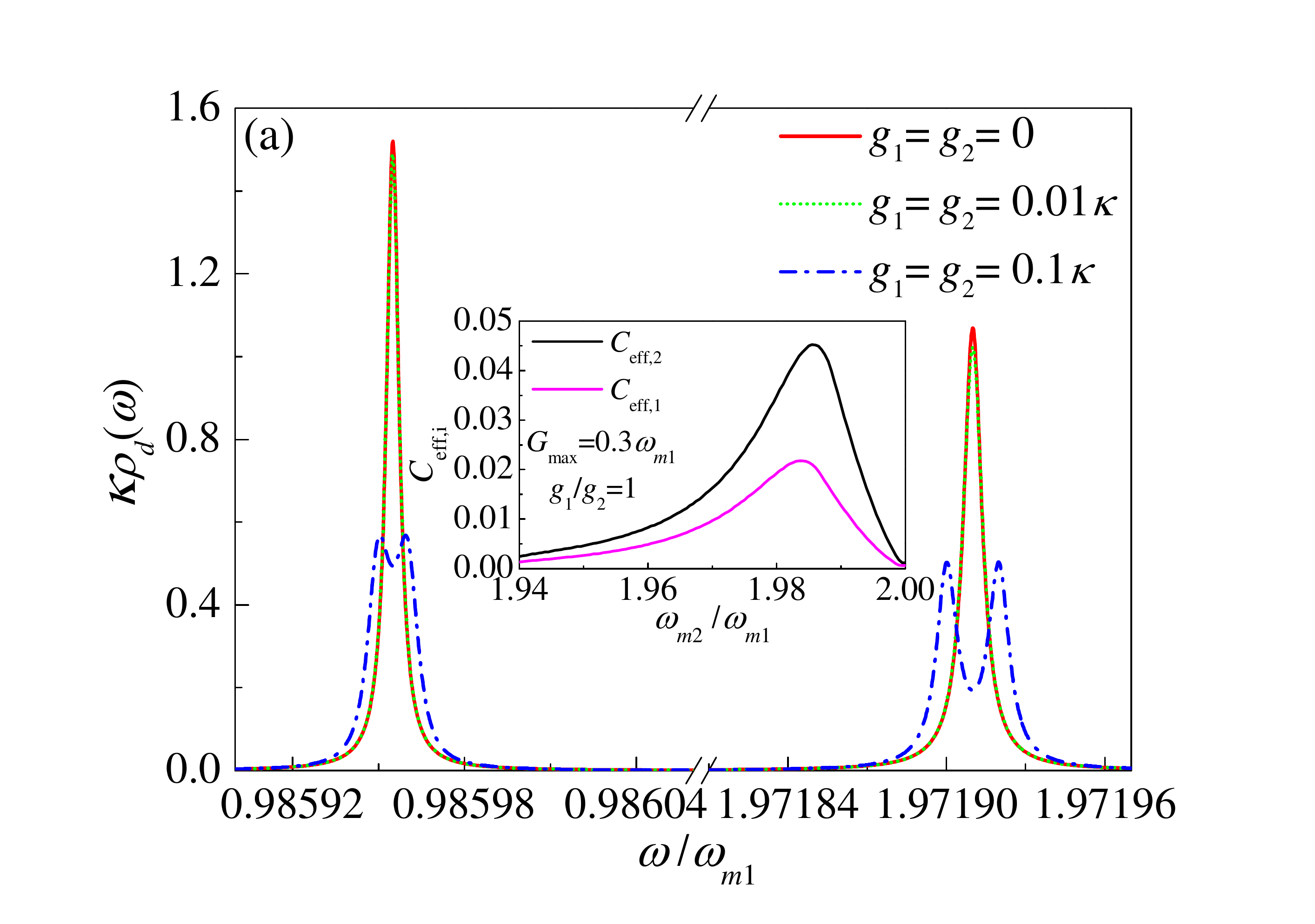}
\includegraphics[clip,width=9cm]{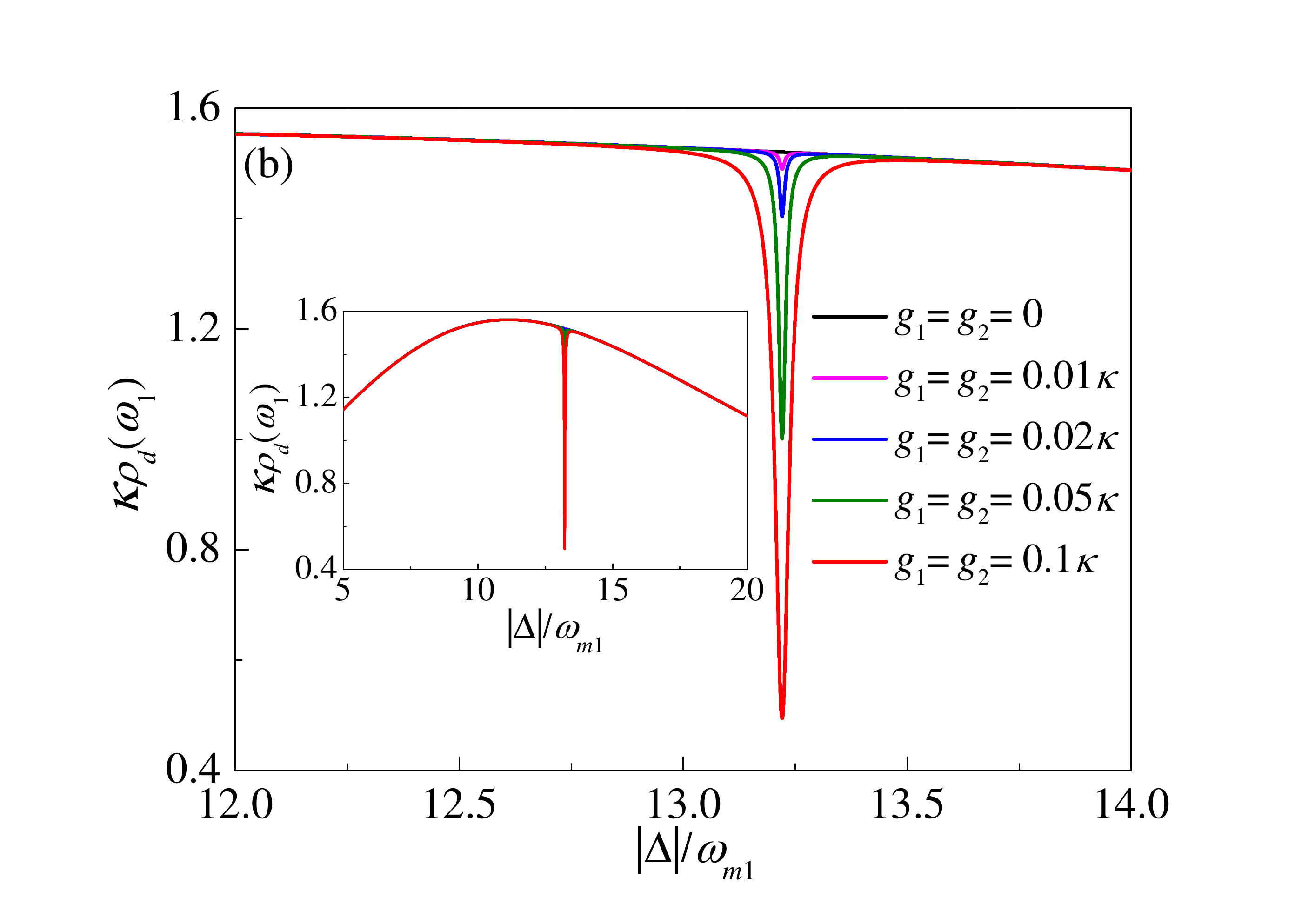}
\caption{(a): Comparison of the cavity DOS in the linear (red solid curve) and nonlinear regimes, using $g_{1,2}=0.01\kappa$ (green dotted curve) and $g_{1,2}=0.1\kappa$ (blue dot-dashed curve). We also used: $|\Delta|= 13.22 \omega_{m1}$, $G_2= 0.3\omega_{m1}$, $\omega_{m2}/\omega_{m1}= 1.9858$, $\kappa=0.02\omega_{m1}$, $\gamma_{1,2}=10^{-4}\kappa$.  In the inset we show a comparison of $C_{\mathrm{eff},1}$ (bottom curve) and  $C_{\mathrm{eff},2}$ (top curve). The two curves are already optimized  over $\Delta$, with $g_{1,2}= 0.01\kappa$. Other parameters are the same of the main plot. (b): Dependence of $\rho_d(\omega_1)$ as function of $\Delta$ for several values of $g_{1,2}$  (other parameters are as in the upper panel). The sharp dip corresponds to the resonant condition and becomes more pronounced at larger values of $g_{1,2}$. The inset shows the same plot in a larger range of $\Delta$.
\label{fig:DOS}}
\end{figure}
%%%%%%%%%%%%%%%%%%%%%%%%%%%%%%%%%%%%%%%%%%%%%%%%%%%%%%

%%%%%%%%%%%%%%%%%%%%%%%%%%%%%%%%%%%%%%%%%%%%%%%%%%%%%%
\subsection{Lineshape and lower polariton}
%%%%%%%%%%%%%%%%%%%%%%%%%%%%%%%%%%%%%%%%%%%%%%%%%%%%%%

We conclude this section by discussing the qualitative change of lineshape induced by nonlinear effects. Since here the damping rates of the two polaritons are comparable [see for example below Eq.~(\ref{kappa_reduced_approx}), where we have obtained $\kappa_2\simeq 2\kappa_1$], the spectral lineshape is not modified qualitatively at small $g$. This behavior is illustrated by the $g=0.01 \kappa$ curves of Fig.~\ref{fig:DOS}(a) and is distinct from what happens in a two-mode system, where a sharp dip can be induced at the higher polariton peak for very small values of $g$  \cite{Lemonde2013PRL, Lemonde2015PRA, Borkje2013PRL}. Therefore, similar to the 4-mode optomechanical ring \cite{Jin2018PRA}, the nonlinear effects could be more easily demonstrated by tuning external parameters like $\Delta$ and $G_i$ across the resonant condition $\omega_2=2\omega_1$. As shown in  Fig.~\ref{fig:DOS}(b), this  will induce a sharp feature in the dependence of the density of states (or a related observable, e.g., the OMIT signal \cite{Lemonde2015PRA, Jin2018PRA}).

Instead, if the optomechanical coupling can be made larger, one enters in a regime where two distinct resonances appear, as illustrated in Fig.~\ref{fig:DOS} by assuming $g_{1,2}=0.1\kappa$. The splittings of the $\omega_1,\omega_2$ polariton peaks are respectively given by:
\begin{equation}
\delta_1  \simeq  4 g_{211}\sqrt{ n_1-n_2}, \qquad
\delta_2  \simeq 4 g_{211}  \sqrt{ n_1+1/2}, \label{splittings}
\end{equation}
and they are resolved when $\delta_{1,2} \gtrsim \kappa_{1,2}$.

Figure~\ref{fig:DOS}(a) also shows that the nonlinear effects at the upper ($\omega_2$) and lower ($\omega_1$) polaritons are comparable. This fact can be checked from our previous analytical expressions: In the regime of negligible $\gamma_{1,2}$ we have $\kappa_2/ \kappa_1 \simeq n_1/n_2 \simeq 2$, leading to $C_{\mathrm{eff,1}} \simeq \frac 2 3 C_{\mathrm{eff,2}}$ [based on Eq.~(\ref{eq:C_eff})]. On the other hand, around the maximum of the effective cooperativity we can estimate $\kappa_2/ \kappa_1 \simeq n_1/n_2 \simeq 3/2$, leading to $C_{\mathrm{eff,1}} \simeq \frac 2 5 C_{\mathrm{eff,2}}$.

%%%%%%%%%%%%%%%%%%%%%%%%%%%%%%%%%%%%%%%%%%%%%%%%%%%%%%%%%%
\section{Conclusion}\label{Sec: summary}
%%%%%%%%%%%%%%%%%%%%%%%%%%%%%%%%%%%%%%%%%%%%%%%%%%%%%%%%%%

In this paper, we investigate the nonlinear interaction effects in a three-mode cavity optomechanical system with one cavity mode and two mechanical modes. To take full the advantage of the two mechanical modes, we concentrate on a regime where  resonant scattering of phonon-like polaritons takes place.

Because of the very small polariton dissipation rates, the nonlinear effects on the cavity density of states and related observables can be greatly enhanced compared to a regular optomechanical system. In the large detuning limit and considering an upper bound on the largest achievable dressed coupling, we obtain the optimal value of the nonlinear effects. Our analytic expressions of the optimal value indicate that the nonlinear effects can be enhanced by a parameter which is typically large, since it is proportional to the ratio $\kappa/\gamma$.

Although with small single-photon optomechanical couplings $g_{1,2}$ the nonlinear effects only induce a slight modification of the spectral lineshape, it would still be possible to observe sharp features by tuning system parameters across the resonant condition. On the other hand, if a regime of sufficiently large $g_{1,2}$ can be reached, the splittings of the $\omega_1,\omega_2$ polariton peaks are clearly established.

\begin{acknowledgments}

S.C. acknowledges support from the National Key Research and Development Program of China (Grant No.~2016YFA0301200), NSAF (Grant No. U1930402), NSFC (Grants No. 11974040 and No. 1171101295), and a Cooperative Program by the Italian Ministry of Foreign Affairs and International Cooperation (No. PGR00960). Y.-D. Wang acknowledges support from NSFC (Grant No. 11947302) and MOST (Grant No. 2017FA0304500). L.J.J. acknowledges support from NSFC (Grant No. 11804020). It is our pleasure to thank helpful discussions with A. A. Clerk.

\end{acknowledgments}

\appendix

%%%%%%%%%%%%%%%%%%%%%%%%%%%%%%%%%%%%%%%%%%%%%%%%%%%%%%%%%%%%%%%
\section{Equal mechanical frequencies}\label{appendix_equal_wm}
%%%%%%%%%%%%%%%%%%%%%%%%%%%%%%%%%%%%%%%%%%%%%%%%%%%%%%%%%%%%%%%

If the two mechanical resonators have the same frequency, i.e., $\omega_{m1}=\omega_{m2}=\omega_{m}$, the system becomes equivalent to a two-mode optomechanical cavity.
This is easily seen by introducing the mechanical dark mode $b_{-}$ and the bright mode $b_{+}$:
\begin{align}
b_{-} & =  \frac{G_{1}b_{2}-G_{2}b_{1}}{\widetilde{G}},\label{eq:dark mode}\\
b_{+} & =  \frac{G_{1}b_{1}+G_{2}b_{2}}{\widetilde{G}},\label{eq:bright mode}
\end{align}
with $\widetilde{G}=\sqrt{G_{1}^{2}+G_{2}^{2}}$. Hence, we rewrite the Hamiltonian as follows:
\begin{align}
H_{l}+H_{nl}  = -\Delta d^{\dagger}d+\omega_{m}b_{+}^{\dagger}b_{+}+\omega_{m}b_{-}^{\dagger}b_{-} \nonumber \\
          +\widetilde{G}(d+d^{\dagger})(b_{+}+b_{+}^{\dagger})+ \widetilde{g}d^{\dagger}d(b_{+}+b_{+}^{\dagger}),
\end{align}
where $\widetilde{g}=\widetilde{G}/\sqrt{N}$. We see that only the bright mode interacts with the cavity and the optomechanical interaction has the standard form.

%%%%%%%%%%%%%%%%%%%%%%%%%%%%%%%%%%%%%%%%%%%%%%%%%%%%%%%%%%
\section{Approximate form of $V$}\label{Appendix_V}
%%%%%%%%%%%%%%%%%%%%%%%%%%%%%%%%%%%%%%%%%%%%%%%%%%%%%%%%%%

In this Appendix, we present the approximate form of $V$ in the large detuning limit $|\Delta| \gg \omega_{mi},G_i$. We first perform a block-diagonalization of $M$ using quasi-degenerate perturbation theory:
\begin{align} \label{M_blocks}
 e^{-S}M e^S  \simeq
\left(\begin{array}{ccc}
\omega_{m1}^2 - B_{11}^2& -B_{12}^2& 0\\
-B_{21}^2 & \omega_{m2}^2- B_{22}^2 & 0\\
0 & 0 & \Delta^2 + B^2_{11} +B^2_{22}
\end{array}\right),
\end{align}
where the $B_{ij}^2$, given by  Eq.~(\ref{B_def}), are the second-order correction with respect to the unperturbed matrix $M_{i,j}^{(0)}=\Delta^2 \delta_{3,i}\delta_{3,j}$. For easier reference, we repeat here their expression:
\begin{equation}\label{B_def_repeated}
B^2_{ij}  =  \frac{4}{|\Delta|}G_{i}G_j \sqrt{\omega_{mi}\omega_{mj}}.
\end{equation}
To lowest order, the transformation matrix $S$ is given by:
\begin{align}
S & \simeq  \frac{1}{|\Delta|} \left(\begin{array}{ccc}
0 & 0 & B_{11} \\
0 & 0 & B_{22} \\
 -B_{11} & -B_{22} & 0 \\
\end{array}\right).
\end{align}
The eigenvalues of Eq.~(\ref{M_blocks}) are the normal mode frequencies $\omega_{i}$ presented in Eq.~(\ref{w12}). Diagonalization of Eq.~(\ref{M_blocks}) is simply through a rotation by an angle $\theta$. Finally, combining $e^S$ and the rotation by $\theta$, we find that $M$ is diagonalized by:
\begin{equation}\label{eq:transformation matrix_1-1}
U  \simeq
\left(\begin{array}{ccc}
\cos\theta & -\sin\theta & \frac{B_{11}}{|\Delta|}\\
\sin\theta & \cos\theta & \frac{B_{22}}{|\Delta|} \\
 -\frac{B_{11}\cos\theta+B_{22}\sin{\theta}}{|\Delta|} & \frac{B_{11}\sin\theta-B_{22}\cos{\theta}}{|\Delta|}  & 1
\end{array}\right),
\end{equation}
where $\tan 2\theta =2 B_{12}^2/(\omega_{m2}^2-\omega_{m1}^2+B_{11}^2-B_{22}^2)$.

The above Eq.~(\ref{eq:transformation matrix_1-1}) can be inserted in Eqs.~(\ref{V_blocks}) and (\ref{Vpm}) to get the desired approximate form of $V$. For example, in the calculation of $\kappa_{i}$ ($i=1,2$) we need the following quantities:
\begin{align}\label{Vmatrix_simplified}
V_{3,i}^2-V_{3,i+3}^2 &= U_{3,i}^2, \nonumber \\
(V_{1,i}+V_{1,i+3})^2 & = U_{1,i}^2 \frac{\omega_{m1}}{\omega_i},\nonumber \\
(V_{2,i}+V_{2,i+3})^2 &= U_{2,i}^2  \frac{\omega_{m2}}{\omega_i},
\end{align}
which are readily obtained from Eq.~(\ref{eq:transformation matrix_1-1}). Furthermore, in the main text we impose the additional restriction $G_i \ll \omega_{m1},\omega_{m2}$. Then, it is not difficult to show that the rotation angle $\theta$ is small. In Eq.~(\ref{Vmatrix_simplified}) we can approximate $U_{1,1}= U_{2,2} =1$ and  $U_{1,2}= U_{2,1} = 0$, giving $\kappa_{1,2}$ as in Eq.~(\ref{pert_theory_kappa}). Equation~(\ref{pert_theory_n_g}) for $n_i,g_{211}$ is obtained in a similar way.

\end{document}